\documentclass[a4paper,twocolumn,amsmath,amssymb,pra,showpacs,showkeys]{revtex4}

\usepackage{graphicx}
\usepackage{dcolumn} 

\graphicspath{{Figures/}}

\newcommand{\magCs}{\ensuremath{{}^{133}\mathrm{Cs}}}

\newcommand{\alignment}{\ensuremath{\mathcal{A}_{0}}}

\newcommand{\TheSlope}{\ensuremath{\mathrm{t_\omega}}}
\newcommand{\NEM}{\ensuremath{\mathrm{NEM}}}
\newcommand{\cplotgrscale}{0.38}
\newcommand{\grscale}{0.44}
\newcommand{\largegrscale}{0.47}
\newcommand{\myvector}{\boldsymbol}
\newcommand{\rf}{\ensuremath{\mathrm{rf}}}
\newcommand{\liaphase}{\ensuremath{\phi_l}}

\newcommand{\Mx}{\ensuremath{\mathrm{M}_x}}

\newcommand{\Hz}{\ensuremath{\mathrm{Hz}}}
\newcommand{\fTHz}{\ensuremath{\mathrm{fT}/\sqrt{\Hz}}}
\newcommand{\uW}{\ensuremath{\mu\mathrm{W}}}
\newcommand{\HzuW}{\ensuremath{\Hz/\uW}}
\newcommand{\mHzuWs}{\ensuremath{\mathrm{m}\Hz/{\uW}^2}}

\newcommand{\T}{\rule{0pt}{2.6ex}}
\newcommand{\B}{\rule[-1.2ex]{0pt}{0pt}}

\newcommand{\sGamma}[1]{\ensuremath{\Gamma_{\kern-0.23em #1}}}
\newcommand{\PL}{\ensuremath{P_{\kern-0.15em L}}}

\begin{document}

\title{Sensitivity of double resonance alignment magnetometers}

\author{Gianni Di~Domenico}
\email{Gianni.DiDomenico@unifr.ch}
\author{Herv\'e~Saudan}
\author{Georg~Bison}
\author{Paul~Knowles}
\author{Antoine~Weis}

\affiliation{%
Physics Department, University of Fribourg, Chemin du Mus\'ee 3,
1700 Fribourg, Switzerland
}%

\date{31 May 2007}

\begin{abstract}
We present an experimental study of the intrinsic magnetometric
sensitivity of an optical/\rf{}-frequency double resonance
magnetometer in which \emph{linearly\/} polarized laser light is
used in the optical pumping and detection processes.  We show that a
semi-empirical model of the magnetometer can be used to describe the
magnetic resonance spectra. Then, we present an efficient method to
predict the optimum operating point of the magnetometer, i.e., the
light power and \rf{}~Rabi frequency providing maximum magnetometric
sensitivity.  Finally, we apply the method to investigate the
evolution of the optimum operating point with temperature. The
method is very efficient to determine relaxation rates and thus
allowed us to determine the three collisional disalignment cross
sections for the components of the alignment tensor. Both first and
second harmonic signals from the magnetometer are considered and
compared.
\end{abstract}

\pacs{32.60.+i, 32.30.Dx, 07.55.Ge, 33.40.+f}

\keywords{atomic alignment, magnetometer, cesium, relaxation cross
sections}

\maketitle

\section{Introduction}
\label{sec:introduction}

Our group develops optically pumped alkali vapor magnetometers (OPM)
for both applied~\cite{georgJOSA} and fundamental~\cite{GroegerNIST}
research.  The diverse requirements of these demanding applications,
in terms of sensitivity, spatial resolution, scalability, and
measurement bandwidth, warrants investigation of new OPM schemes. An
interesting and promising avenue is the use of atomic {\itshape
alignment\/} instead of {\itshape orientation\/} to probe the
external magnetic field.  We will refer to an OPM based on atomic
alignment as DRAM (double resonance alignment magnetometer), while
we will speak of DROM (double resonance orientation magnetometer)
when the magnetization has the symmetry of an atomic orientation.
Recently, our group presented both theoretical~\cite{DRAMtheo} and
experimental~\cite{DRAMexp} investigations of the magnetic resonance
spectra produced in a cesium vapor in which an alignment is created
and detected by a single linearly--polarized laser beam.

Of direct importance for us, the DRAM scheme has a more flexible
geometry than the well-known DROM \Mx--configuration~\cite{blo62}.
For maximal sensitivity, the DROM scheme requires a 45~deg.~angle
between the laser beam and the magnetic field~\cite{blo62}, limiting
applications calling for a compact arrangement of multiple sensors.
In multichannel devices, as required for cardiomagnetic
measurements~\cite{georg2} for example, the DRAM method offers the
advantage that the laser beam can be oriented either parallel or
perpendicular to the offset field without loss of sensitivity.

The line shapes of the second harmonic DRAM signal have significantly
narrower linewidths than the DROM signal under identical conditions.
Narrow linewidths suppress systematic errors in optical magnetometers,
visible as long term baseline drifts, and potentially increase the
magnetometric sensitivity.  This means that a DRAM could lead to a
higher magnetometric sensitivity than a DROM for equal signal to noise
ratio.

Moreover, in DROM devices the interaction of the atoms with the
circularly polarized laser light leads to an $M$ dependent energy
shift of the Zeeman hyperfine components when the laser frequency is
not centered on the optical resonance line, the so-called light
shift~\cite{Happer:1967:EOF,Barrat:1961:EPO}. In that case, the
effect of laser power and frequency changes is indistinguishable
from the effect of magnetic field changes, thus limiting the
magnetometric performance and introducing systematic uncertainties
on the determination of the absolute value of the field. In the
DRAM, the linearly polarized light produces a light shift depending
on $M^2$, which does not have the same characteristics as a magnetic
field Zeeman interaction. The $M^2$ shift broadens the magnetic
resonance line, thereby slightly reducing the magnetic sensitivity,
but it will not shift the resonance frequency. The absence of linear
$M$-dependent systematic resonance shifts make the DRAM an
attractive magnetometer for precision experiments searching for
$M$-dependent effects, such as electric dipole moment
searches~\cite{BudkerWeisRMP}.

In this article, we present an experimental study of the
magnetometric sensitivity of a double resonance alignment
magnetometer. The principle of the DRAM with its theoretical
description is given in Sec.~\ref{sec:DRAMprinciple}, the
experimental setup is described in Sec.~\ref{sec:experiment}, and
the operational definition of the magnetometric sensitivity is
introduced in Sec.~\ref{sec:sensitivity}\@.  Then, in
Sec.~\ref{sec:DRAMextension} we show that a simple empirical
extension of the DRAM model extends its validity to significantly
higher laser powers.  This extended model is used in
Sec.~\ref{sec:DRAMoptimum}, where we develop a method to predict the
optimum operating point of the DRAM based on physical parameters
extracted from specific measurements.  Finally, the method is
applied in Sec.~\ref{sec:temperature} to determine the temperature
dependence of the DRAM optimum operating point, and the results
obtained for different cells are compared in
Sec.~\ref{sec:discussion}\@.

\begin{figure}[t]
\includegraphics[width=\largegrscale\textwidth, clip]{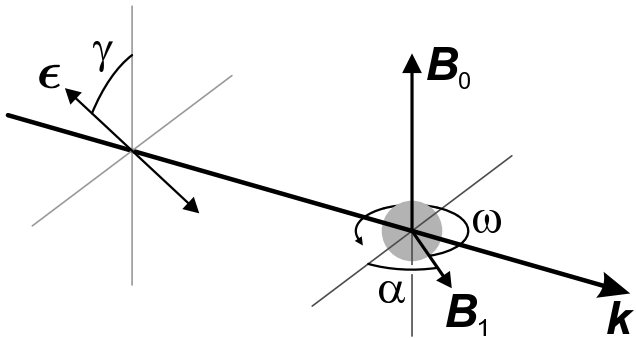}
\caption{Double resonance magnetometer geometry using linearly
polarized light.  Here, $\myvector{k}$ is the direction of the
linearly polarized laser beam.  The \rf{} field $\myvector{B}_{1}$
(shown here at $t=0$) rotates in a plane perpendicular to the static
field $\myvector{B}_{0}$.  The linear polarization vector
$\myvector{\epsilon}$ makes an angle $\gamma$ with the static field
$\myvector{B}_{0}$, and the phase of $\myvector{B}_{1}$ is
characterized by $\alpha$.}
\label{fig:geometry}
\end{figure}

\section{Double resonance alignment magnetometer}
\label{sec:DRAMprinciple}

The geometry of a double resonance alignment magnetometer is
presented in Fig.~\ref{fig:geometry}: it is identical to the one
described in~\cite{DRAMtheo}.  A linearly polarized laser beam, with
polarization $\myvector{\epsilon}$ inclined at angle $\gamma$ to the
magnetic field to be measured, $\myvector{B}_{0}$, is used to create
an atomic alignment via optical pumping in a room temperature vapor
of cesium atoms.  This alignment precesses under the simultaneous
action of the static magnetic field $\myvector{B}_{0}$ and a much
weaker magnetic field $\myvector{B}_{1}$, called the \rf{} field,
rotating at frequency $\omega$ in the plane perpendicular to
$\myvector{B}_{0}$ and driving the magnetic resonance transitions.
Competition between relaxation, optical pumping, and magnetic
resonance produces a steady state in the rotating frame.

The precession of the alignment generates modulations of the
absorption coefficient which create signals at both the fundamental
($\omega$) and the second harmonic ($2\omega$) of the applied \rf{}
frequency $\omega$.  The magnetic resonance signals $S_{\omega}(t)$
and $S_{2\omega}(t)$ are obtained here by monitoring the transmitted
light power with a photodiode. The details of the calculation of
$S_{\omega}(t)$ and $S_{2\omega}(t)$ are given in~\cite{DRAMtheo},
therefore only the most relevant equations needed for the
magnetometric analysis are reproduced here.

The magnetic resonance signals can be written as
\begin{subequations}
\label{eq:signals}
\begin{eqnarray}
  S_{\omega }(t)&\!\!\!=
  \phantom{_2} \alignment\, h_{\omega}(\gamma) \left[ \right.&
    \!\!\! \phantom{-}D_{\omega }(\delta)\,\cos \left(\omega t-\alpha\right)\nonumber \\
  && \!\!\! \left. -  A_{\omega }(\delta)\,\sin \left(\omega
  t-\alpha\right)\right]
  \, ,
\label{subeq:somega}\\
  S_{2\omega }(t)&\!\!\!= \alignment\, h_{2\omega }(\gamma )\left[\right. &
  \!\!\!           -A_{2\omega }(\delta)\,\cos \left(2\omega t-2\alpha\right)\nonumber \\
  && \!\!\! \left. -D_{2\omega }(\delta)\,\sin \left(2\omega
  t-2\alpha\right)\right]
  \,, \quad \phantom{,}
  \label{subeq:s2omega}
\end{eqnarray}
\end{subequations}
where $\alignment$ is the alignment, defined in~\cite{DRAMtheo},
produced by the optical pumping. The angular dependencies of the
first and second harmonic signals are given by
\begin{subequations}
\label{eq:angdep}
\begin{eqnarray}
h_{\omega }(\gamma )
&=&\frac{3}{16}\left( 2\sin 2\gamma +3\sin 4\gamma \right)\,,
\label{subeq:hgamma1} \\
h_{2\omega }(\gamma )
&=&\frac{3}{32}(1-4\cos 2\gamma +3\cos 4\gamma )\,,
\label{subeq:hgamma2}
\end{eqnarray}
\end{subequations}
where $\gamma$ is the angle between the light polarization and the
static field $\myvector{B}_0$.
%
The first and second harmonic signals have both absorptive,
$A_{\omega}(\delta)$, $A_{2\omega }(\delta)$, and dispersive,
$D_{\omega}(\delta)$, $D_{2\omega }(\delta)$, components in their
line shapes, given by
\begin{subequations}
\label{eq:lineshapes3gamma}
\begin{eqnarray}
D_{\omega}(\delta)& = &\frac{\delta\, \sGamma{0}\omega_1
                     (\sGamma{2}^2+4 \delta^2-2 \omega_1^2)}{Z(\delta)}
\, ,
\label{subeq:3gd1}\\
A_{\omega}(\delta)& = &\frac{\phantom{\delta\,}\sGamma{0}\omega_1
                    \left[(\sGamma{2}^2 + 4 \delta^2) \sGamma{1}
                         + \sGamma{2} \omega_1^2 \right]} {Z(\delta)}
\, ,
\label{subeq:3ga1}\\
D_{2\omega}(\delta)& = &\frac{\delta\, \sGamma{0} \omega_1^2
                      (2\sGamma{1}+\sGamma{2})}{Z(\delta)}
\, ,
\label{subeq:3gd2}\\
A_{2\omega}(\delta)& = & \frac{\phantom{\delta\,}\sGamma{0}
                       \omega_1^2 (\sGamma{1}\sGamma{2} - 2\delta^2
                       +\omega_1^2)}{Z(\delta)}
\, ,
\label{subeq:3ga2}
\end{eqnarray}
\end{subequations}
with a resonance denominator,
\begin{eqnarray}
Z(\delta)&=& \sGamma{0} \left(\sGamma{1}^2+\delta^2\right)
\left(\sGamma{2}^2
     + 4\delta^2 \right) \nonumber \\
 & &{}+\left[\sGamma{1} \sGamma{2} \left(2\sGamma{0} + 3\sGamma{2}\right)
     - 4 \delta ^2 \left(\sGamma{0} - 3\sGamma{1} \right) \right] \omega_1^2
      \nonumber \\
 & &{}+ \left(\sGamma{0} + 3\sGamma{2}\right) \omega_1^4 \,.
\label{eq:denom3gamma}
\end{eqnarray}
In Eqs.~(\ref{eq:lineshapes3gamma}) and~(\ref{eq:denom3gamma}),
$\omega_{1}=\gamma_{F}B_{1}$ is the Rabi frequency of the \rf{} field
where $\gamma_{F}$ is the gyromagnetic ratio of the ground state
hyperfine level $F$.  The detuning $\delta=\omega-\omega_{0}$ is the
difference between the \rf{} frequency $\omega$ and the Larmor
frequency $\omega_{0}=\gamma_{F}B_{0}$, and $\sGamma{0}$,
$\sGamma{1}$, $\sGamma{2}$ are alignment relaxation rates.  More
precisely, the DRAM model~\cite{DRAMtheo} calculates the evolution of
the alignment multipole moments $m_{2,q}$ via a density matrix
approach (for a general introduction to the use of multipole moments
in the density matrix formalism, see~\cite{blum}). The moments
$m_{2,q}$ are defined with respect to a quantization axis aligned with
$\myvector{B}_0$ and relax with rates $\sGamma{|q|}$. In practice,
both the absorptive $A_{\omega}(\delta)$, $A_{2\omega }(\delta)$, and
dispersive, $D_{\omega}(\delta)$, $D_{2\omega }(\delta)$, components
of the signals can be used to measure the magnetic field, and can be
isolated by phase-sensitive detection of the transmission signals
$S_{\omega }(t)$ and $S_{2\omega }(t)$.

The effect of $\gamma$, the angle between the linear polarization
vector and the static magnetic field, is contained in the functions
$h_{\omega }(\gamma)$ and $h_{2\omega }(\gamma)$.  The first harmonic
signal is maximized for $\gamma=25.5\deg$ and the second harmonic
signal for $\gamma=90\deg$.  Thus we distinguish between two
realizations of the DRAM:
\begin{enumerate}
  \item the first harmonic DRAM, choosing $\gamma=25.5\deg$ and
  measuring $A_{\omega}(\delta)$, $D_{\omega}(\delta)$, and
  \item the second harmonic DRAM, choosing $\gamma=90\deg$ and
  measuring $A_{2\omega}(\delta)$, $D_{2\omega}(\delta)$.
\end{enumerate}
Since the line shapes are different, we expect the two realizations of
the DRAM to result in distinct optimum operating points and
magnetometric sensitivities.

\section{Experimental setup}
\label{sec:experiment}

\begin{figure}
\centerline{\includegraphics*[width=\grscale\textwidth]{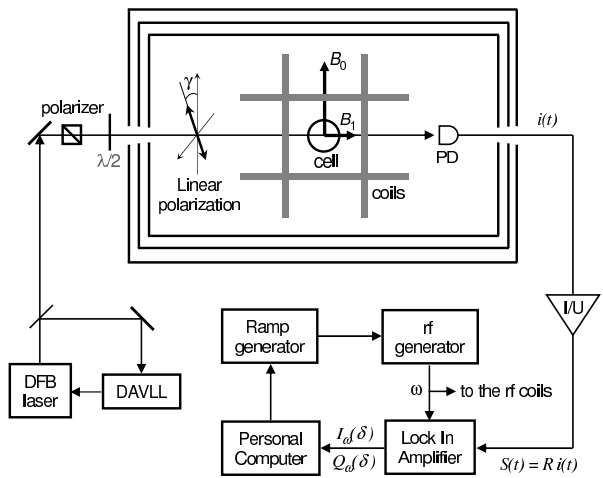}}
\caption{Experimental setup: A cell containing Cs vapor was mounted
inside a 3-axis Helmholtz coil array all placed inside a three-layer
mu-metal shield.  The polarization angle $\gamma$, measured with
respect to the offset field $\myvector{B}_{0}$, was set by a ~linear
polarizer followed by a half-wave plate ($\lambda /2$) located
outside the shield for ease of access.  The laser light traversed
the cell and was converted to a current by a nonmagnetic photodiode
(PD).  All details can be found in the text.} \label{fig:expsetup}
\end{figure}

The experimental setup used for the optimization procedure is shown in
Fig.~\ref{fig:expsetup}.  A~Pyrex cell, paraffin-coated for spin
relaxation suppression and evacuated except for an atomic cesium vapor
in thermal equilibrium with a metal droplet, provided the paramagnetic
atom sample.  The cell was isolated from ambient magnetic fields by a
three-layer mu-metal shield (ID$\;300$~mm, length $580$~mm,
OD$\;=590$~mm).  Inside the shield, a primary pair of Helmholtz coils
produced a static magnetic field $\myvector{B}_{0}$ of about
$3\:\mu\mathrm{T}$ perpendicular to the light propagation direction.
Additional orthogonal pairs of Helmholtz coils (only one pair is shown
in Fig.~\ref{fig:expsetup}) were used to suppress residual fields and
gradients.
An \rf{} magnetic field $\myvector{B}_{1}$, rotating at approximately
10~kHz in the plane perpendicular to the static magnetic field, was
created by a set of two pairs of Helmholtz coils, wound on the same
supports as the static field coils.  All internal structural
components were made from nonmagnetic materials.

The laser beam used to pump and probe the atomic vapor confined in the
cell was generated by a distributed feedback (DFB) diode laser, with a
wavelength of $894~\mbox{nm}$, stabilized to the $6S_{1/2},
F_g\!=\!4\!\rightarrow\!6P_{1/2}, F_e\!=\!3$ hyperfine transition by
means of a dichroic atomic vapor laser lock
(DAVLL)~\cite{WiemanDAVLL}.  A~linear polarizer followed by a
half-wave plate prepared linearly polarized light of adjustable
orientation $\gamma$ with respect to $\myvector{B}_{0}$.  The residual
circular polarization contamination was measured to be less than 1\%.
A~nonmagnetic photodiode, followed by a low-noise transimpedance
amplifier, detected the light power transmitted through the cell.  The
resulting signal was analyzed by a lock-in amplifier tuned either to
the first or second harmonic of the \rf{} frequency, depending on the
DRAM configuration under study (cf.~Sec.~\ref{sec:DRAMprinciple}).
A~computer recorded magnetic resonance spectra by initiating the \rf{}
frequency sweep and simultaneously recording the in-phase and
quadrature signals from the lock-in amplifier.  The computer also
controlled the light and \rf{} power delivered to the cell, and
measured the total light power on the photodiode as well as the
temperature of the apparatus.  The system was thus automated and could
make measurements of the magnetic resonance signals for ranges of
light and \rf{} powers.  Forced air heating was used to make
temperature changes to the system, changes that were slow with respect
to the time needed to record one spectrum.

In practice, lock-in detection of the signals given by
Eqs.~(\ref{eq:signals}) with respect to the \rf{}~frequency $\omega$ adds
a phase \liaphase{} (selectable in the lock-in amplifier) and a small
pick-up signal $p_{(1,2)(A,D)}$ (smaller than 1\% of the signal at
maximum) to each of the line shapes given by
Eqs.~(\ref{eq:lineshapes3gamma})~\cite{DRAMexp}.  Due to \liaphase,
the in-phase and quadrature spectra are, in general, a mixture of
dispersive and absorptive line shapes.  Demodulation of the signal
Eq.~(\ref{subeq:somega}), yields expressions used to fit the recorded
in-phase and quadrature spectra
\begin{subequations}
\label{eq:demod_one}
\begin{eqnarray}
I_\omega(\delta)\!
   &\! = g_{\omega}(\PL) h_{\omega}(\gamma) &
    \kern-1ex
    \left[ \phantom{+}(D_{\omega}(\delta)+p_{1D})\cos{(\alpha\!+\!\liaphase)}
    \right.\qquad \nonumber  \\
   &   & \kern-1ex \left.
           \, +(A_{\omega}(\delta)+p_{1A})\sin{(\alpha\!+\!\liaphase)}
         \right] ,
\label{subeq:demod_ip} \\
Q_\omega(\delta)\!
   &\! = g_{\omega}(\PL) h_{\omega}(\gamma) &
    \kern-1ex
    \left[ \phantom{+}(A_{\omega}(\delta)+p_{1A})\cos{(\alpha\!+\!\liaphase)}
    \right. \nonumber  \\
   &   & \kern-1ex \left.
           \, -(D_{\omega}(\delta)+p_{1D})\sin{(\alpha\!+\!\liaphase)}
         \right] .
\label{subeq:demod_qu}
\end{eqnarray}
\end{subequations}
The $g_{\omega}(\PL)$ factor is used here to contain not only
amplifier gain factors, but also the alignment $\alignment$ and any
light power, $\PL$, dependencies.  A similar mix of
$A_{2\omega}(\delta)$ and $D_{2\omega}(\delta)$ was used for the
second harmonic signal given by Eq.~(\ref{subeq:s2omega}).

\begin{figure}[t]
\centerline{\includegraphics[width=\grscale\textwidth]{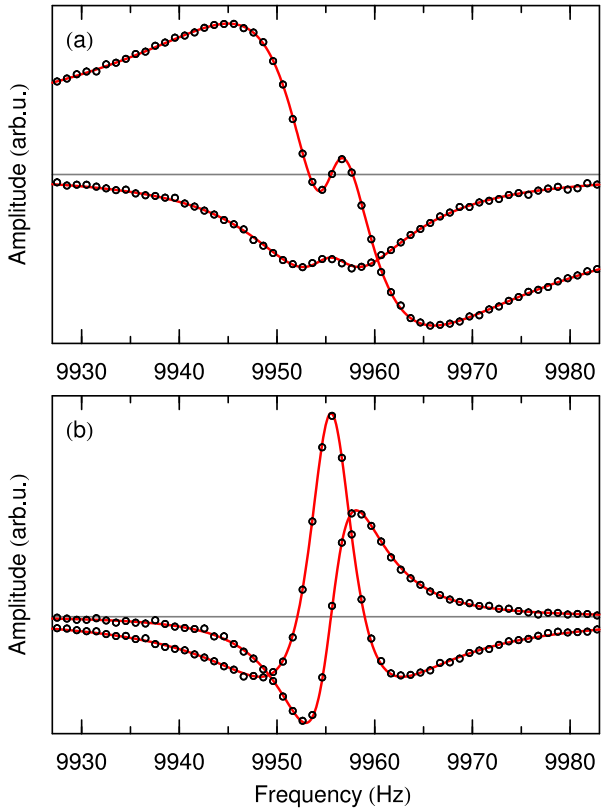}}
\caption{(Color online) Measurements (circles) of the in-phase and
quadrature magnetic resonance signals detected as amplitude
modulations of the transmitted light power.  (a) First harmonic
signals.  (b) Second harmonic signals.  The statistical uncertainty
on each data point is represented by the symbol size.  The solid
lines are fits of the theoretical line shapes given by
Eqs.~(\protect\ref{eq:lineshapes3gamma})--(\protect\ref{eq:demod_one}).}
\label{fig:DRAMsignals}
\end{figure}

Typical measured signals for the first and second harmonic magnetic
resonance spectra are presented in Fig.~\ref{fig:DRAMsignals},
together with fits of the theoretical line shapes given by
Eqs.~(\ref{eq:lineshapes3gamma})--(\ref{eq:demod_one}).  All four
curves are fitted simultaneously with one set of relaxation rates; for
the presented data $\sGamma{0}=2\pi\times 1.64(2)\:\Hz$,
$\sGamma{1}=2\pi\times 2.93(2)\:\Hz$ and $\sGamma{2}=2\pi\times
3.08(2)\:\Hz$.  Detailed information on the fitting procedure is found
in~\cite{DRAMexp}.  The excellent quality of the fits allow us to
extract the amplitude $g$, the relaxation rates $\sGamma{i}$, and the
Rabi frequency $\omega_1$ from a single set of double resonance
spectra.  For that reason no calibration of the \rf-coils is needed.
This is an advantage compared to the DROM where the Rabi
frequency and the longitudinal relaxation rate are correlated to the
point where they cannot be individually extracted from measured line
shapes.

\section{Magnetometric sensitivity}
\label{sec:sensitivity}

The dispersive magnetic resonance line shapes given by
Eqs.~(\ref{subeq:3gd1}) and~(\ref{subeq:3gd2}) have a linear
dependence on the detuning $\delta=\omega - \gamma_F B_{0}$ at the
center of the resonance.  By proper choice of \liaphase{}, the quadrature
signal, Eq.~(\ref{subeq:demod_qu}), can be made completely dispersive,
giving direct access to the linear zero crossing of the resonance,
\begin{equation}
Q_{\omega}(B_{0}) = g_{\omega}(\PL) h_{\omega}(\gamma)\;
D_{\omega}(\omega - \gamma_F B_{0})\,,
\label{eq:Qomega}
\end{equation}
with a similar expression for the $2\omega$ resonance.
At constant $\omega$, $Q_\omega$ can be used as a magnetometer
signal for a limited range of magnetic field strengths $|\omega -
\gamma_F B_{0}| \ll \Gamma$, where $\Gamma$ is the resonance
linewidth.
In that range, a change of $B_0$ by a small $\Delta B_0$ can be
measured as a change of $Q_\omega$ by the amount
\begin{equation}
\Delta Q_\omega  = \Delta B_0 \left. \frac{dQ_\omega}{dB_0}
\right|_{B_0=\omega/\gamma_F}  = \Delta B_0\, \TheSlope\,.
\label{eq:deltaq}
\end{equation}
The slope $\TheSlope$ is obtained from fits of the dispersive
experimental magnetic resonance line shape
(Fig.~\ref{fig:DRAMsignals}) using the relation
\begin{equation}
\TheSlope =  \left. \frac{dQ_\omega}{dB_0} \right|_{B_0=\omega/\gamma_F}
  = \frac{-1}{\gamma_F} \left.\frac{dQ_\omega}{d \delta} \right|_{\delta=0}
\,.
\label{eq:slope}
\end{equation}
Again, similar relations were used for the $2\omega$ signals.

The noise equivalent magnetic field, or \NEM, represents the noise
limit on the derived value of $B$ given the noise in $Q$.  The total
noise in $Q$ has contributions from external magnetic field
fluctuations and from all sources of technical noise, like laser
intensity and frequency noise (converted to intensity noise by the
atomic vapor), electronic noise, and so on.  All technical noises
can, in principle, be reduced until the system reaches the fundamental
limit arising from the photocurrent shot noise.  Therefore, we use
the shot noise limited \NEM{} as the measure for comparing
the performance of different magnetometric schemes~\cite{georgJOSA}.

The root spectral density of the photocurrent shot noise is given
by
\begin{equation}
\rho_S = R\,\sqrt{2 e I_{DC}}\,,
\label{eq:noise}
\end{equation}
where $R$ is the transimpedance gain of the current amplifier, $e$
the electron charge, and $I_{DC}$ the DC photocurrent.
Given $\rho_S$, the \NEM{} can be expressed as a root spectral
density of field fluctuations by inverting Eq.~(\ref{eq:deltaq})
\begin{equation}
\NEM = \frac{\rho_S}{ |\TheSlope|} \,.
\label{eq:NEM}
\end{equation}
Since $\rho_S$ was evaluated from a measurement of the
photocurrent before the lock-in amplifier, the internal gain
correction of the lock-in was used to give a measurement of
$\TheSlope$ usable in Eq.~(\ref{eq:NEM}).

The goal of this study was to find the optimum DRAM operating
parameters, $\PL$ and $\omega_1$, yielding maximum magnetometric
sensitivity, i.e., minimal \NEM.

\section{Empirical extension of the DRAM model}
\label{sec:DRAMextension}

As discussed in~\cite{DRAMtheo,DRAMexp}, the analytical expressions
for the DRAM model [Eqs.~(\ref{eq:signals})--(\ref{eq:denom3gamma})]
are valid for low laser power only, however, empirical formulas were
presented modeling the light power dependence of the relaxation rates
and of the global amplitude factors of the DRAM signals.  Here, we
present improved empirical formulas extending the DRAM model to even
higher light powers, our goal being to cover the power domain that
must be explored while optimizing the magnetometer.

The following empirical formula successfully represents the laser
power dependence of the first harmonic signal
\begin{equation}
    g_{\omega}(\PL) = C
    \frac{\PL^2}{\left(P_{S1}+\PL\right)\left(P_{S2}+\PL\right)}
    \label{eq:Aext}
\end{equation}
where $C$ is a constant and $P_{S1}$, $P_{S2}$ are experimentally
determined saturation powers for which we currently have no rigorous
model in terms of fundamental physical constants and processes.  A
similar formula applies to the second harmonic amplitude, but requires
different values for both the constant and the saturation powers.  The
model reflects the expectation that both the creation of alignment as
well as the ability to probe the alignment will saturate with
increasing power.

\begin{figure}[t]
\centerline{\includegraphics[width=\grscale\textwidth]{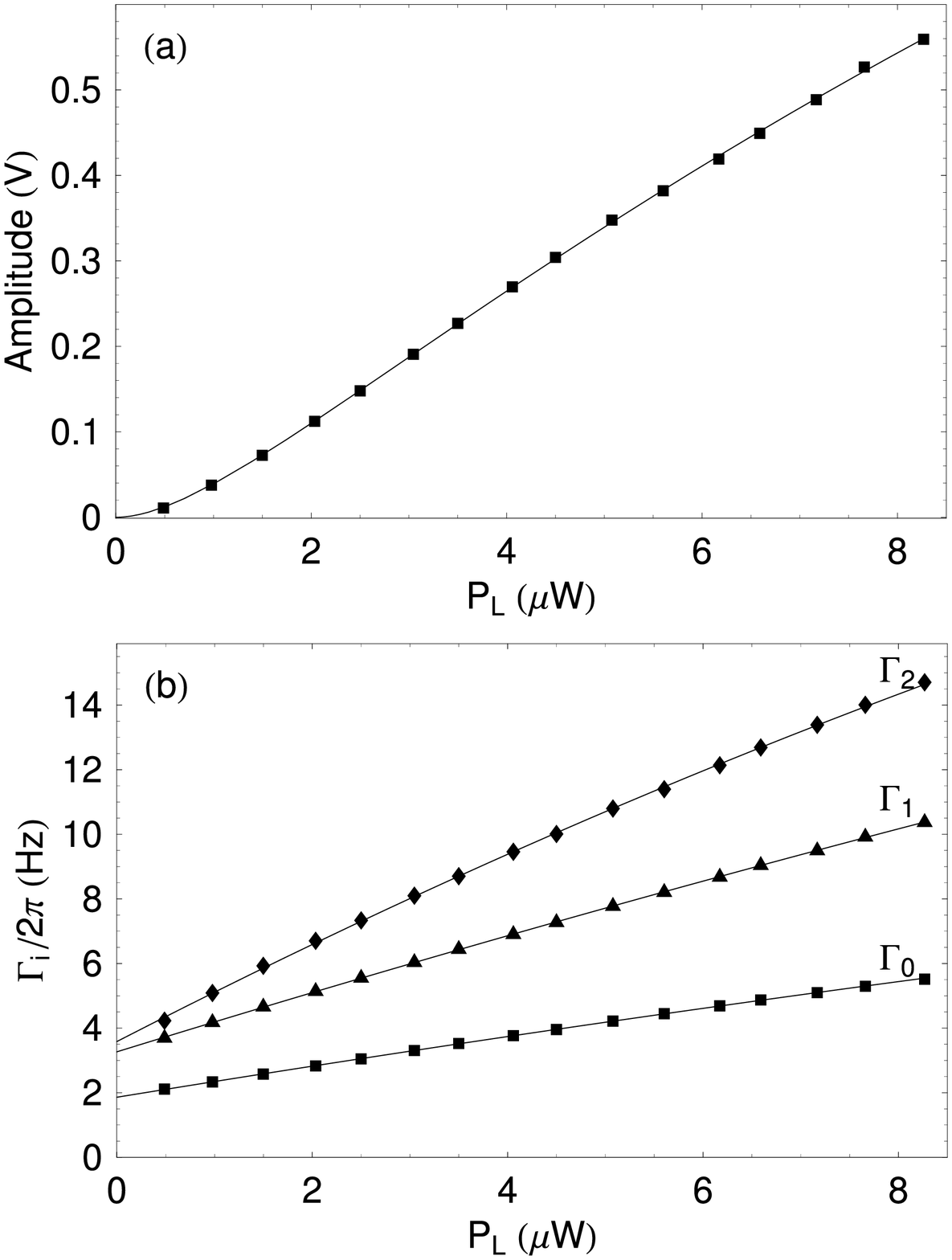}}
\caption{(a) First harmonic DRAM signal amplitude as a function of
laser power.  (b) Relaxation rates as a function of laser power.
Points are measured values, extracted from the fit of the DRAM model
[Eqs.~(\protect\ref{eq:lineshapes3gamma})] to the experimental
magnetic resonance spectra.  The solid lines are fits of the
extended DRAM model [Eqs.~(\protect\ref{eq:Aext})
and~(\protect\ref{eq:G1G2G3ext})] to the experimental data.  The
data were measured at room temperature from a first harmonic DRAM
with $\gamma=25.5\deg$, $\omega_1=2\pi\times 8.3\:\Hz$.}
\label{fig:1wPlot}
\end{figure}

\begin{figure}[t]
\centerline{\includegraphics[width=\grscale\textwidth]{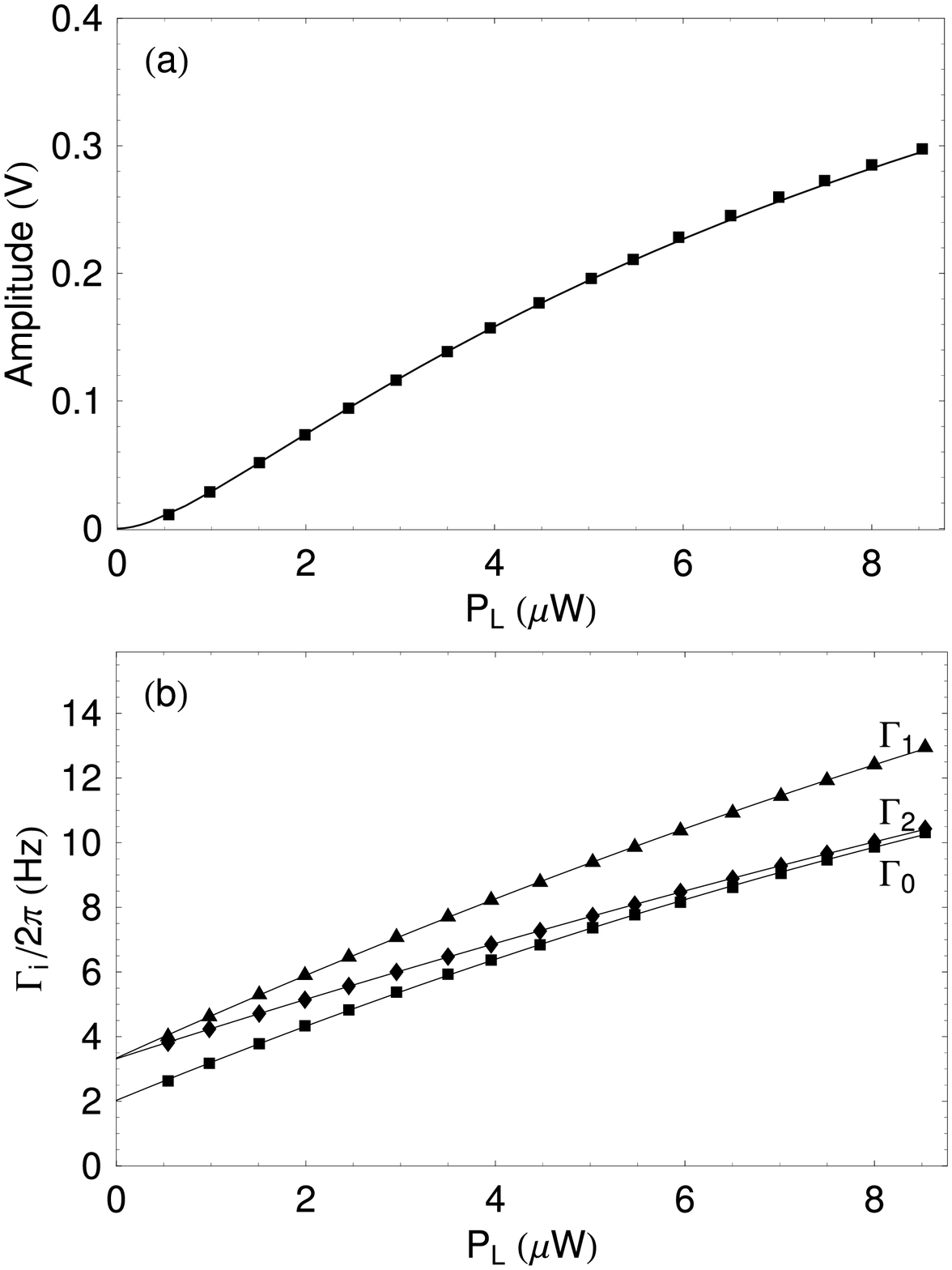}}
\caption{(a) Second harmonic DRAM signal amplitude as a function of
laser power.  (b) Relaxation rates as a function of laser power.
Points are measured values extracted from the fit of the DRAM model
[Eqs.~(\protect\ref{eq:lineshapes3gamma})] to the experimental
magnetic resonance spectra. The solid lines are fits of the extended
DRAM model [Eqs.~(\protect\ref{eq:Aext})
and~(\protect\ref{eq:G1G2G3ext})] to the experimental data.  The
data were measured at room temperature from a second harmonic DRAM
with $\gamma=90\deg$, $\omega_1=2\pi\times 8.3\:\Hz$.}
\label{fig:2wPlot}
\end{figure}

In a similar way, the $\PL$ dependence of the relaxation rates has
been modeled by a power series and good agreement with the measured
data was found using a second order polynomial for each rate
\begin{subequations}
\label{eq:G1G2G3ext}
\begin{eqnarray}
    \sGamma{0}(\PL) &=& \sGamma{00} + \alpha_{0} \PL + \beta_{0} \PL^{2}\,,
    \label{subeq:G0ext}\\
    \sGamma{1}(\PL) &=& \sGamma{10} + \alpha_{1} \PL + \beta_{1} \PL^{2}\,,
    \label{subeq:G1ext}\\
    \sGamma{2}(\PL) &=& \sGamma{20} + \alpha_{2} \PL + \beta_{2} \PL^{2}\,.
    \label{subeq:G2ext}
\end{eqnarray}
\end{subequations}
We call the following parameter set the {\em extended DRAM model
parameters},
\begin{equation}
\left\{C, P_{S1}, P_{S2}, \sGamma{00},\alpha_{0},\beta_{0}, \sGamma{10},\alpha_{1},\beta_{1},\sGamma{20},\alpha_{2},\beta_{2}\right\}
\end{equation}
and note that they have to be determined experimentally.  For that
purpose, we have measured a series of double resonance spectra as a
function of laser power, and extracted the amplitude and relaxation
rates from the simultaneous fits, using common parameters, of the
theoretical line shapes given by
Eqs.~(\ref{eq:lineshapes3gamma})--(\ref{eq:demod_one}) to the
experimental data, as explained in~\cite{DRAMexp}.  The measurements
were made separately for the first harmonic, with $\gamma=25.5\deg$,
and for the second harmonic, with $\gamma=90\deg$.  The results are
presented in Fig.~\ref{fig:1wPlot}(\ref{fig:2wPlot}) for the
first(second) harmonic signals.

For determining $C$, $P_{S1}$, and $P_{S2}$, the
empirical model of Eq.~(\ref{eq:Aext}) was fitted to the amplitude
data, and the resulting fits are displayed as solid lines in the upper
graphs of Figs.~\ref{fig:1wPlot} and~\ref{fig:2wPlot}.  To~find the
remaining extended parameters, $\sGamma{00}$, $\sGamma{10}$,
$\sGamma{20}$, $\alpha_{0}$, $\alpha_{1}$, $\alpha_{2}$, $\beta_{0}$,
$\beta_{1}$, and $\beta_{2}$, the empirical model of
Eqs.~(\ref{eq:G1G2G3ext}) was fitted to the $\PL$ dependence of the
measured relaxation rates, and the resulting fits are displayed as
solid lines in the lower graphs of Figs.~\ref{fig:1wPlot}
and~\ref{fig:2wPlot}.  Clearly, the extended model accurately
represents the data over the whole range of light powers investigated.

Table~\ref{tab:coeff} summarizes the extended DRAM model parameters
for both the first and the second harmonic signals.  The expectation,
based on the cylindrical symmetry of the physical system,
that $\sGamma{10}$ should equal $\sGamma{20}$ is not reflected in the
data, but the discussion of this will be delayed until
Sec.~\ref{sec:temperature}\@.

\begin{table}[t]
\caption{Experimental values of the extended DRAM model parameters
extracted from the fits to the experimental data presented in
Figs.~\protect\ref{fig:1wPlot} and \protect\ref{fig:2wPlot}.  Only
statistical uncertainties are shown.  See text for details.}
\label{tab:coeff}
\centerline{
\begin{tabular}{|c|c|c|}
\hline
\T \B  Fit parameters & $\gamma=25.5\deg$ & $\gamma=90\deg$  \\ \hline \hline
\T $C$                & $3.1(3)$ V        & $1.0(2)$ V       \\
\T $P_{S1}$           & $1.7(2)$ \uW      & $1.2(2)$ \uW     \\
\T $P_{S2}$           & $24(3)$  \uW      & $11(3)$ \uW      \\
\hline
\T $\sGamma{00}/2\pi$ & $1.86(1)$ \Hz      & $2.03(3)$ \Hz     \\
\T $\sGamma{10}/2\pi$ & $3.27(1)$ \Hz      & $3.34(2)$ \Hz     \\
\T $\sGamma{20}/2\pi$ & $3.58(5)$ \Hz      & $3.31(1)$ \Hz     \\
\hline
\T $\alpha_{0}/2\pi$  & $0.492(4)$ \HzuW    & $1.21(1)$ \HzuW   \\
\T $\alpha_{1}/2\pi$  & $0.934(7)$ \HzuW    & $1.33(1)$ \HzuW   \\
\T $\alpha_{2}/2\pi$  & $1.55(3)$ \HzuW    & $0.946(4)$ \HzuW   \\
\hline
\T $\beta_{0}/2\pi$   & $-5.6(5)$ \mHzuWs   & $-28(2)$ \mHzuWs \\
\T $\beta_{1}/2\pi$   & $-8.9(9)$ \mHzuWs   & $-24(1)$ \mHzuWs \\
\T $\beta_{2}/2\pi$   & $-26(3)$ \mHzuWs  & $-13.4(5)$ \mHzuWs \\
\hline
\end{tabular}
}
\end{table}

\section{Prediction of the DRAM optimum operating point}
\label{sec:DRAMoptimum}

\subsection{Description of the method}
\label{sec:DRAMoptimumtheo}

The optimum operating point of a DRAM can be predicted
from the measured extended DRAM parameters presented in
the previous section.
%
Here, the optimum operating point refers to the laser power
$\PL$ and Rabi frequency $\omega_1$ which minimize the
intrinsic \NEM{} defined in Eq.~(\ref{eq:NEM}).
In that equation, the photocurrent shot noise $\rho_s$ is calculated
from the DC photocurrent using Eq.~(\ref{eq:noise}), and the
on-resonance slope of the magnetometer signal is calculated from the
derivative of the dispersive component of the resonance spectra, see
Eqs.~(\ref{eq:Qomega}) and~(\ref{eq:slope}).
By direct differentiation, we obtain
\begin{equation}
    \left.\frac{dQ_{\omega}}{d\delta}\right|_{\delta=0}
    = \frac{g_{\omega}(\PL) h_{\omega}(\gamma)\,
            \sGamma{0}(\sGamma{2}^2-2\omega_1^2) \omega_1}
    {\left(\sGamma{1}\sGamma{2}+\omega_1^2\right)
     \left[\sGamma{0}\sGamma{1}\sGamma{2}+
                 (\sGamma{0}+3\sGamma{2})\,\omega_1^2\right]}
\label{eq:slopeatresonance1}
\end{equation}
for the slope of the first harmonic signal, and
\begin{equation}
    \left.\frac{dQ_{2\omega}}{d\delta}\right|_{\delta=0}
    = \frac{g_{2\omega}(\PL) h_{2\omega}(\gamma)\,
            \sGamma{0} (2\sGamma{1}+\sGamma{2}) \omega_1^2}
    {\left(\sGamma{1}\sGamma{2}+\omega_1^2\right)
     \left[\sGamma{0}\sGamma{1}\sGamma{2}+
                  (\sGamma{0}+3\sGamma{2})\,\omega_1^2\right]}
\label{eq:slopeatresonance2}
\end{equation}
for the slope of the second harmonic signal.  Combining the above with
the power scaling model of Eqs.~(\ref{eq:Aext})
and~(\ref{eq:G1G2G3ext}) and using the result in Eq.~(\ref{eq:NEM}),
the intrinsic \NEM{} as a function of laser power $\PL$ and Rabi
frequency $\omega_1$ is found.

\begin{figure}[t]
\begin{tabular}{l}
\centerline{\includegraphics[width=\cplotgrscale\textwidth]{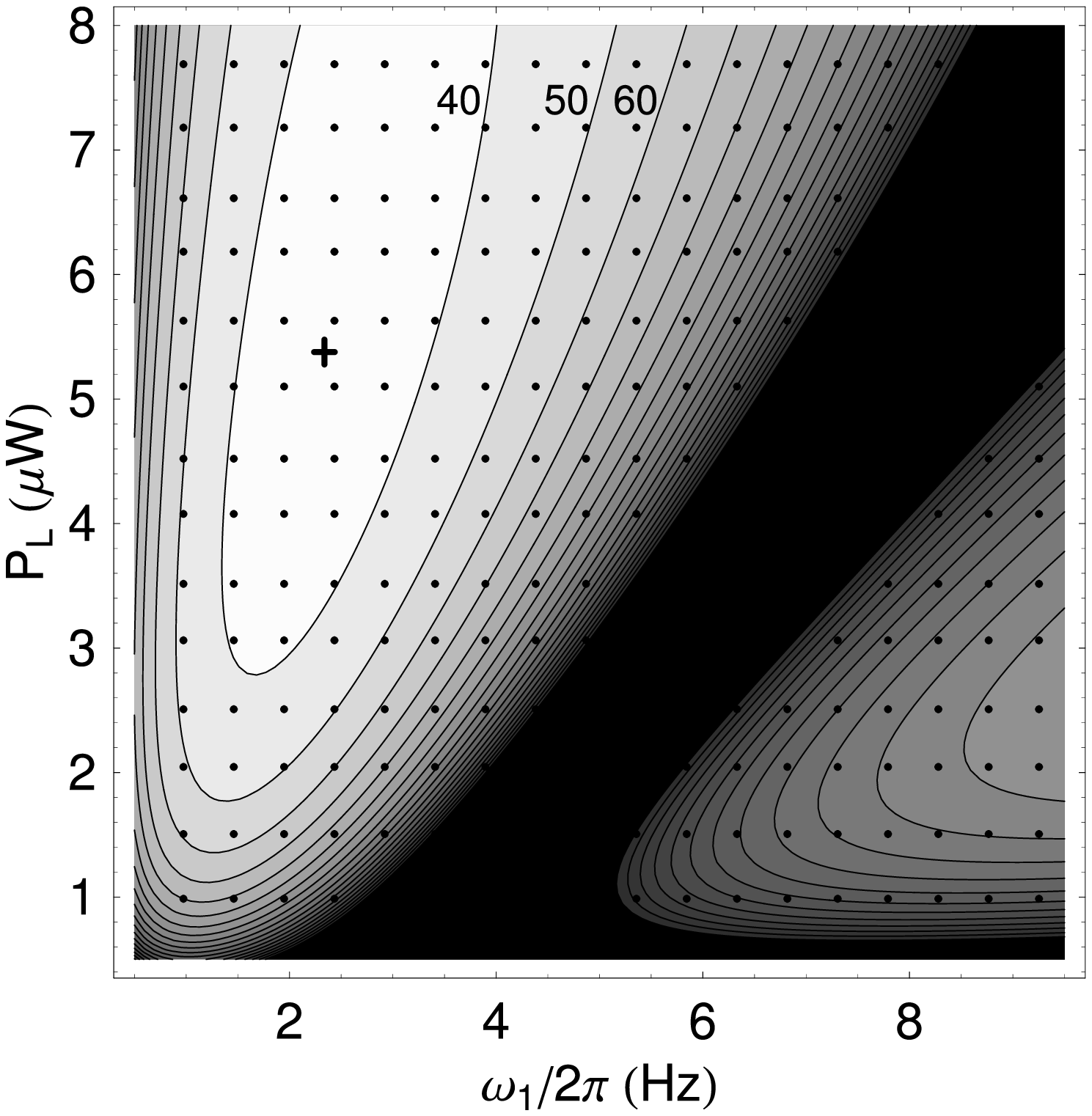}}\\
\centerline{\includegraphics[width=\cplotgrscale\textwidth]{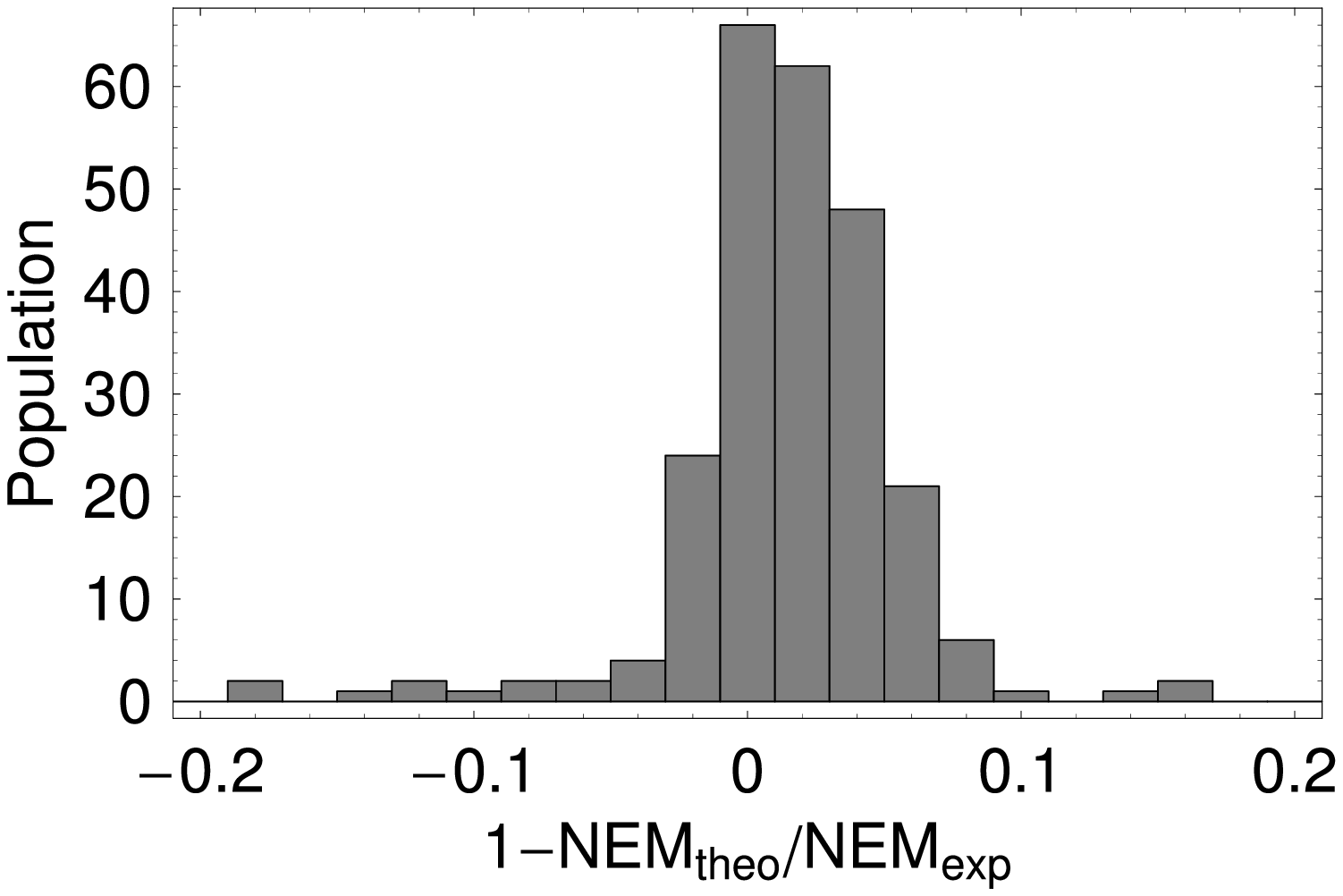}} \\
\end{tabular}
\caption{First harmonic DRAM, $\gamma=25.5\deg$.  The upper graph is a
contour plot of the \NEM{} as a function of $\PL$ and $\omega_1/2\pi$.
The \NEM{} values were calculated using the method developed in
Sec.~\protect\ref{sec:DRAMoptimumtheo}.  The cross indicates the
position where the \NEM{} is minimum (see
Table~\protect\ref{tab:TEoptimum}).  The contour lines start at
$40~\fTHz$ and are spaced by $10~\fTHz$.  The dots indicate the points
in parameter space where the \NEM{} has been measured,
cf.~Sec.~\ref{sec:DRAMoptimumexp}.  The distribution of the relative
difference between calculated and measured values is shown in the
lower graph.}
\label{fig:1wContourPlot}
\end{figure}

\begin{figure}[t]
\begin{tabular}{l}
\centerline{\includegraphics[width=\cplotgrscale\textwidth]{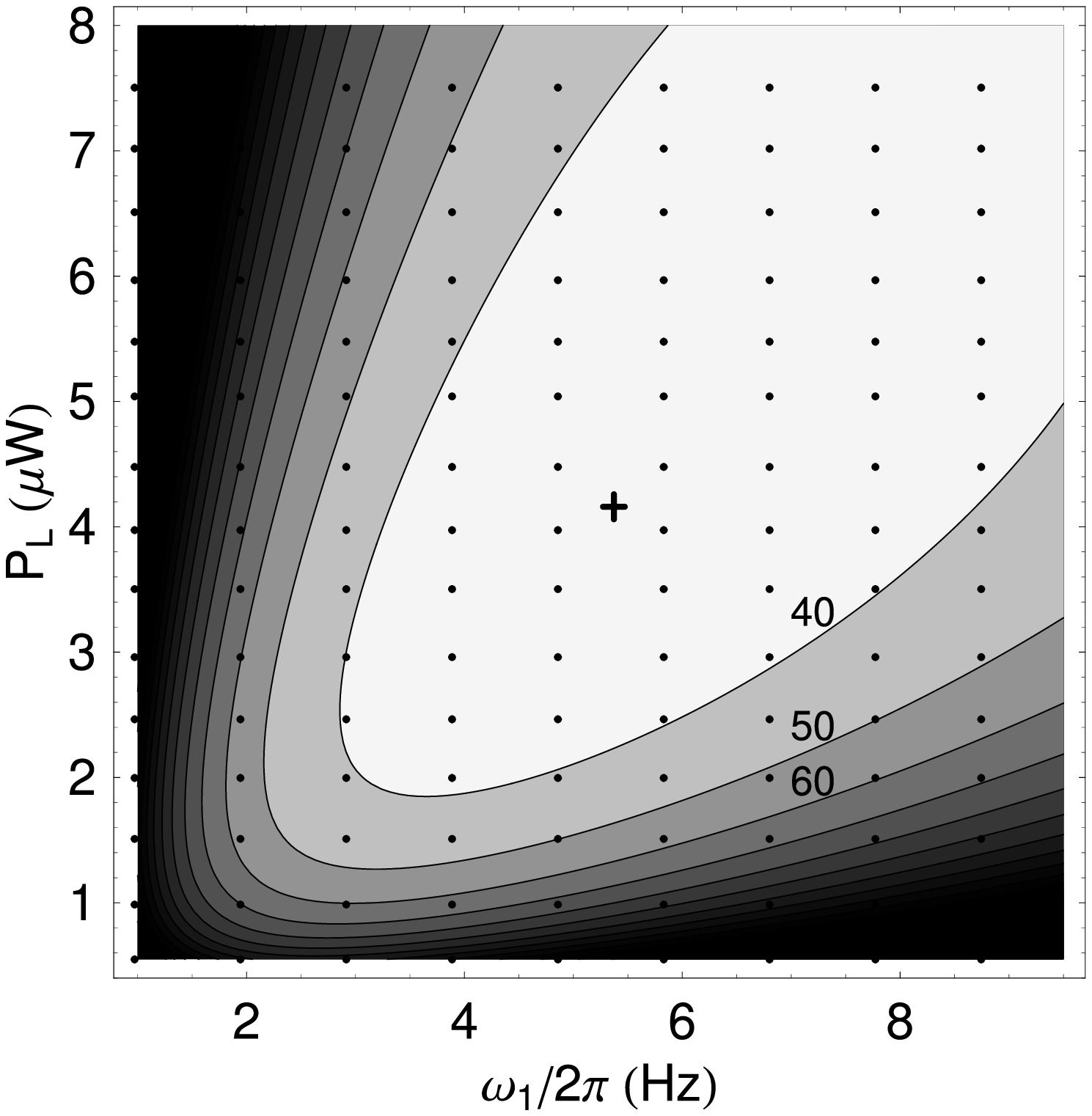}}\\
\centerline{\includegraphics[width=\cplotgrscale\textwidth]{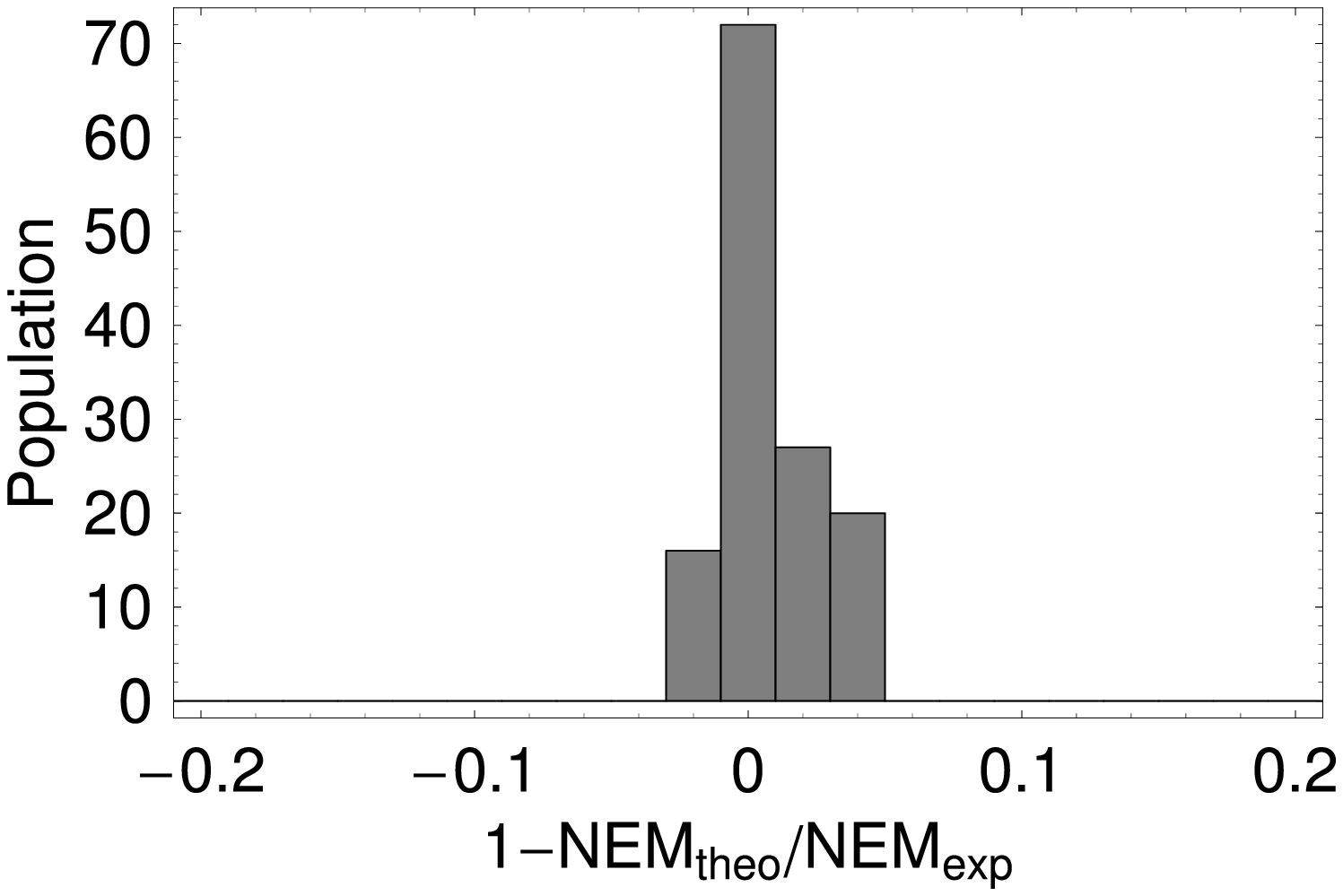}} \\
\end{tabular}
\caption{Second harmonic DRAM, $\gamma=90\deg$.  The upper graph is a
contour plot of the \NEM{} as a function of $\PL$ and $\omega_1/2\pi$.
The \NEM{} values were calculated using the method developed in
Sec.~\protect\ref{sec:DRAMoptimumtheo}.  The cross indicates the
position where the \NEM{} is minimum (see
Table~\protect\ref{tab:TEoptimum}).  The contour lines start at
$40~\fTHz$ and are spaced by $10~\fTHz$.  The dots indicate the points
in parameter space where the \NEM{} has been measured,
cf.~Sec.~\protect\ref{sec:DRAMoptimumexp}.  The distribution of the
relative difference between calculated and measured values is shown in
the lower graph.}
\label{fig:2wContourPlot}
\end{figure}

This \NEM{} function has been calculated for the extended DRAM model
parameters given in Table~\ref{tab:coeff}\@.  The resulting contour
plots of \NEM{} as a function of $\PL$ and $\omega_1$ are presented in
the upper graph of Fig.~\ref{fig:1wContourPlot} for the first harmonic
DRAM, and in the upper graph of Fig.~\ref{fig:2wContourPlot} for the
second harmonic DRAM.  Both graphs show a clear optimum point where
the \NEM{} is minimum.  Table~\ref{tab:TEoptimum} lists the
coordinates of these optimum points, together with the corresponding
\NEM{} value.

\begin{table}
\caption{Theory predictions of laser power $\PL$ and \rf{} field
Rabi frequency $\omega_1$ minimizing the \NEM{} compared to the
experimental best values.  The calculations used the empirical
extension of the DRAM model from
Sec.~\protect\ref{sec:DRAMextension}.} \label{tab:TEoptimum}
\newcolumntype{f}[1]{D{.}{.}{#1}}
\centerline{
\begin{tabular}{|c|f{6}|f{8}|f{4}|} \hline
 \T DRAM            &\multicolumn{1}{c|}{ Optimum $\PL$}
                                   &\multicolumn{1}{c|}{Optimum $\omega_1/2\pi$}
                                                       &\multicolumn{1}{c|}{ \NEM{}} \\
 \B scheme          &\multicolumn{1}{c|}{(\uW)}
                                   &\multicolumn{1}{c|}{(\Hz)}
                                                       &\multicolumn{1}{c|}{(\fTHz)} \\ \hline\hline
 \T Theo.~$1\omega$ & 5.4        & 2.34                 & 35.5     \\
 \T Expt.~$1\omega$ & 5.1(2)     & 2.40(5)              & 35.7(7)  \\ \hline
 \T Theo.~$2\omega$ & 4.2        & 5.4                  & 32.8     \\
 \T Expt.~$2\omega$ & 4.5(2)     & 5.6(1)               & 32.6(6)  \\ \hline
\end{tabular}
}
\end{table}

For the first harmonic DRAM, Fig.~\ref{fig:1wContourPlot} shows a
diagonal valley along $\omega_{1}=\sGamma{2}(\PL)/\sqrt{2}$ where
the \NEM{} is maximized (i.e., poor sensitivity).  There, the \NEM{}
goes to infinity due to the onset of the narrow spectral feature
(discussed in detail in~\cite{DRAMtheo}) appearing on the dispersive
component of the resonance spectra,
cf.~Fig.~\ref{fig:DRAMsignals}.a), reducing the slope to zero.

\subsection{Verification of the method}
\label{sec:DRAMoptimumexp}

The apparatus described in Sec.~\ref{sec:experiment} was used to
measure the on-resonance slope of the dispersive magnetic resonance
signal.  Then, that slope was inserted in Eq.~(\ref{eq:NEM}) to
determine the experimental value of the intrinsic \NEM.  We repeated
the measurement on a $18\times 14$ grid of $\omega_1$ and $\PL$ values
for the first harmonic DRAM, and a $9\times 15$ value grid for the
second harmonic DRAM\@. These measured points are shown as
dots in the upper graphs of Figs.~\ref{fig:1wContourPlot}
and~\ref{fig:2wContourPlot}.  For all measured points, the difference
between the \NEM{} predicted from the extended model and the measured
value has been determined, and the distribution of the relative
difference is plotted in the lower graph of
Fig.~\ref{fig:1wContourPlot} for the first harmonic DRAM, and in the
lower graph of Fig.~\ref{fig:2wContourPlot} for the second harmonic
DRAM.  Within the experimental uncertainty, there are no significant
differences between the measurements and the predictions.

The experimental optimum operating points, where the measured \NEM{}
is minimized, was found, and the results are shown in
Table~\ref{tab:TEoptimum}\@.  Note that the optimal laser power is
nearly the same for both first and second harmonic DRAMs.  The Rabi
frequency required to optimize the $2\omega$ \NEM{} is more than twice
that of the $1\omega$ {\NEM}.  The minimum \NEM{} is slightly lower
for the second harmonic signal.  Table~\ref{tab:TEoptimum} compares
the measured values with the predicted values calculated using the
extended DRAM model.  The agreement is very good, in particular for
the \NEM{} values.  This means that given the relaxation rates and
saturation powers, the optimum point can be predicted with precision
of 5\% using the extended DRAM model.

\subsection{Advantage of the method}

The automated experimental determination of the optimum operating
point of a DRAM, for a given temperature, can take several tens of
hours.  Indeed, that was the case for the \NEM{} measurements over the
grid of $\PL$ and $\omega_1$ values presented above.

By contrast, to find the optimum operating point based on the
prediction of the extended DRAM model parameters requires only
measurements as a function of $\PL$, since the $\omega_1$ dependence
of the magnetic resonance spectra is perfectly described by the DRAM
model presented in~\cite{DRAMtheo}.  Thus, the measurement time needed
for finding the optimum can be reduced by one order of magnitude when
using the above method to predict the optimum operating point instead
of exploring the whole bidimensional parameter space.  This is
particularly useful to make a rapid characterizing of the quality of
coated Cs cells, and when studying the optimum point as a function of
temperature, the topic of the next section.

\section{DRAM optimum point evolution with temperature}
\label{sec:temperature}

The atomic vapor density and atom velocity distribution (and hence
the interatomic and wall collision rates), as well as the relaxation
probability during individual wall collisions, depend on
temperature. The alignment relaxation rates, $\sGamma{0}$,
$\sGamma{1}$, and $\sGamma{2}$, depend in a nontrivial way on all
those parameters because of three main contributing processes,
namely collisional spin-exchange, wall-collision electron-spin
randomization, and the reservoir
effect~\cite{Bou66,Liberman,Budker:2005:MTN}.
Temperature thus has an important influence on the magnetometric
sensitivity.  Unfortunately, the influence is hard to model and so is
worth measuring.

\begin{figure}[t]
\centerline{\includegraphics[width=\grscale\textwidth]{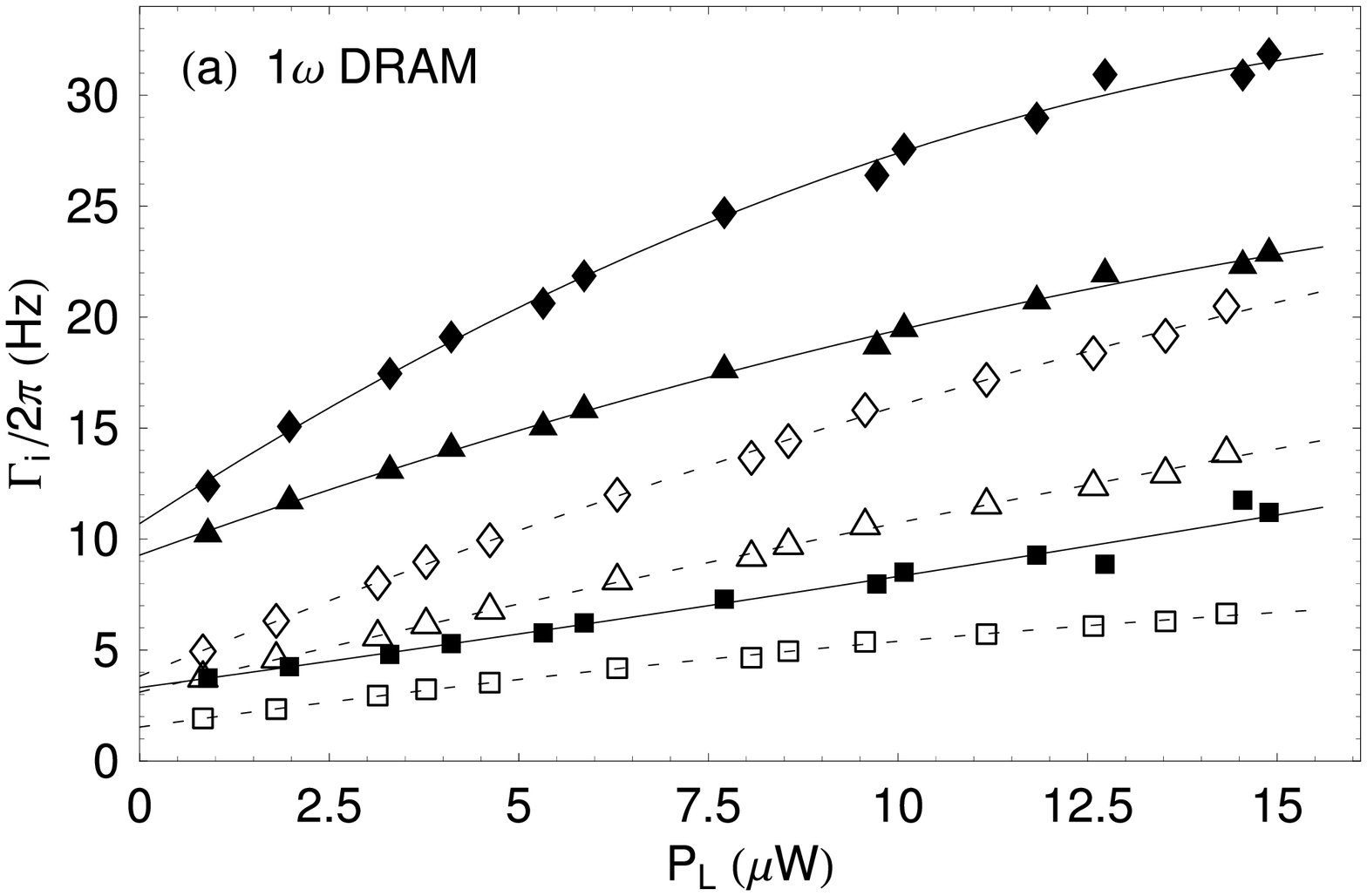}}
\centerline{\includegraphics[width=\grscale\textwidth]{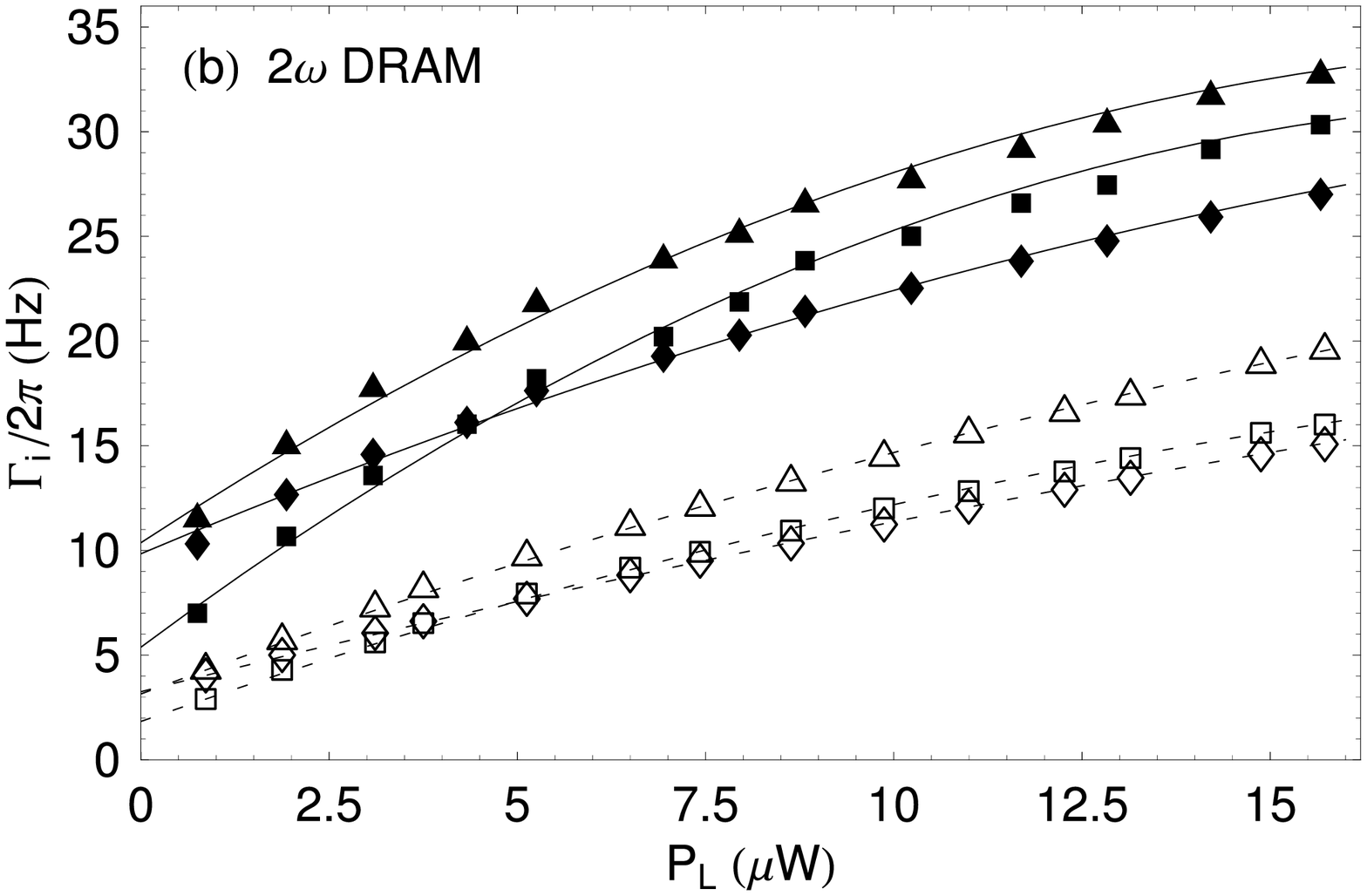}}
\caption{Relaxation rates as a function of laser power, measured at
two different temperatures. The empty symbols
($\square$,$\triangle$,$\lozenge$) represent
$\sGamma{0}$,$\sGamma{1}$,$\sGamma{2}$ measured at $T=25^\circ$C.
The filled symbols ($\blacksquare$,$\blacktriangle$,$\blacklozenge$)
represent $\sGamma{0}$,$\sGamma{1}$,$\sGamma{2}$ measured at
$T=38^\circ$C. The solid and dashed lines are a fit of the extended
DRAM model [Eqs.~(\ref{eq:G1G2G3ext})] to the experimental data. (a)
First harmonic DRAM with $\gamma=25.5\deg$. (b) Second harmonic DRAM
with $\gamma=90\deg$.} \label{fig:GTPlot}
\end{figure}

\begin{figure}[t]
\centerline{\includegraphics[width=\grscale\textwidth]{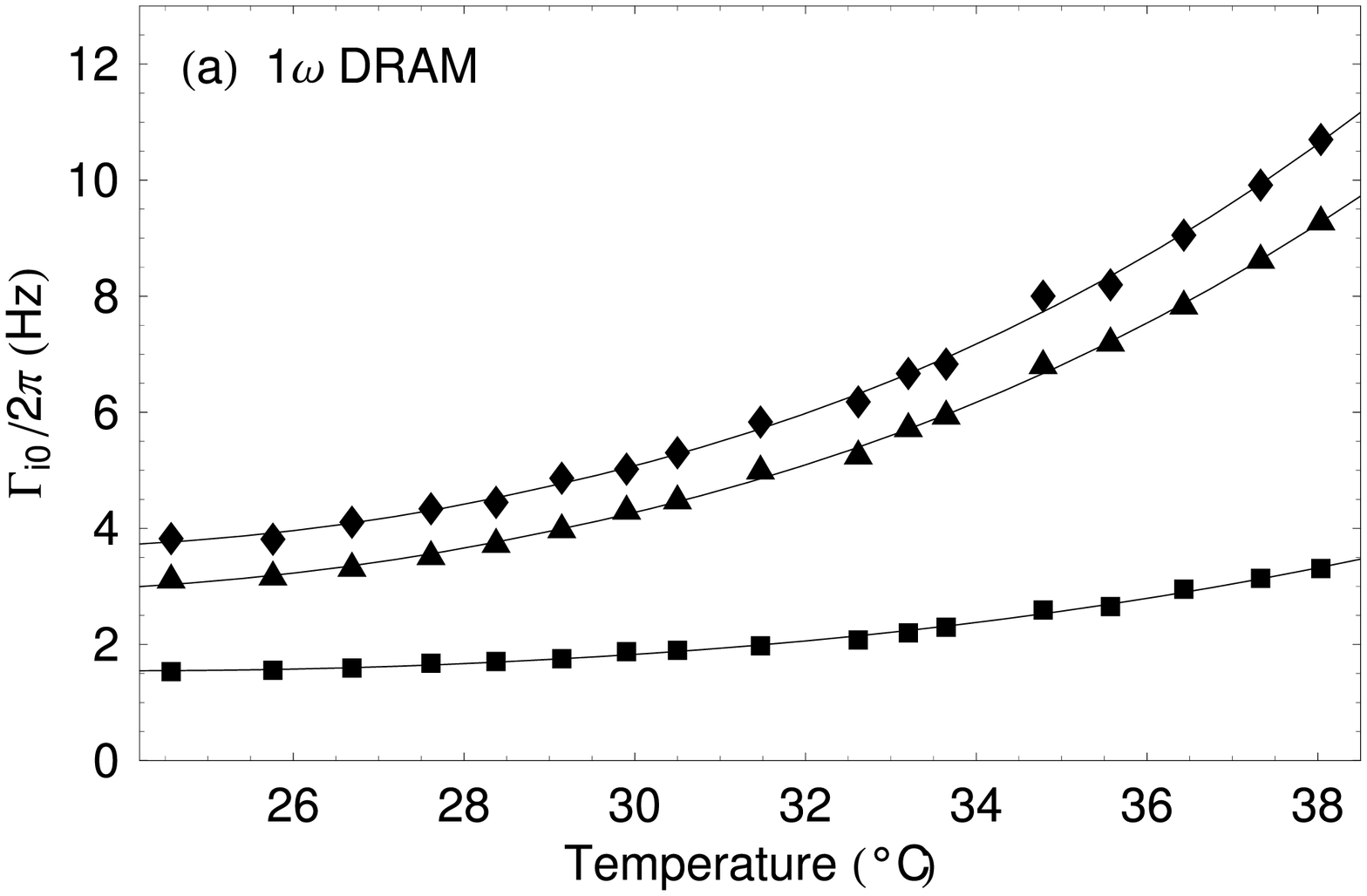}}
\centerline{\includegraphics[width=\grscale\textwidth]{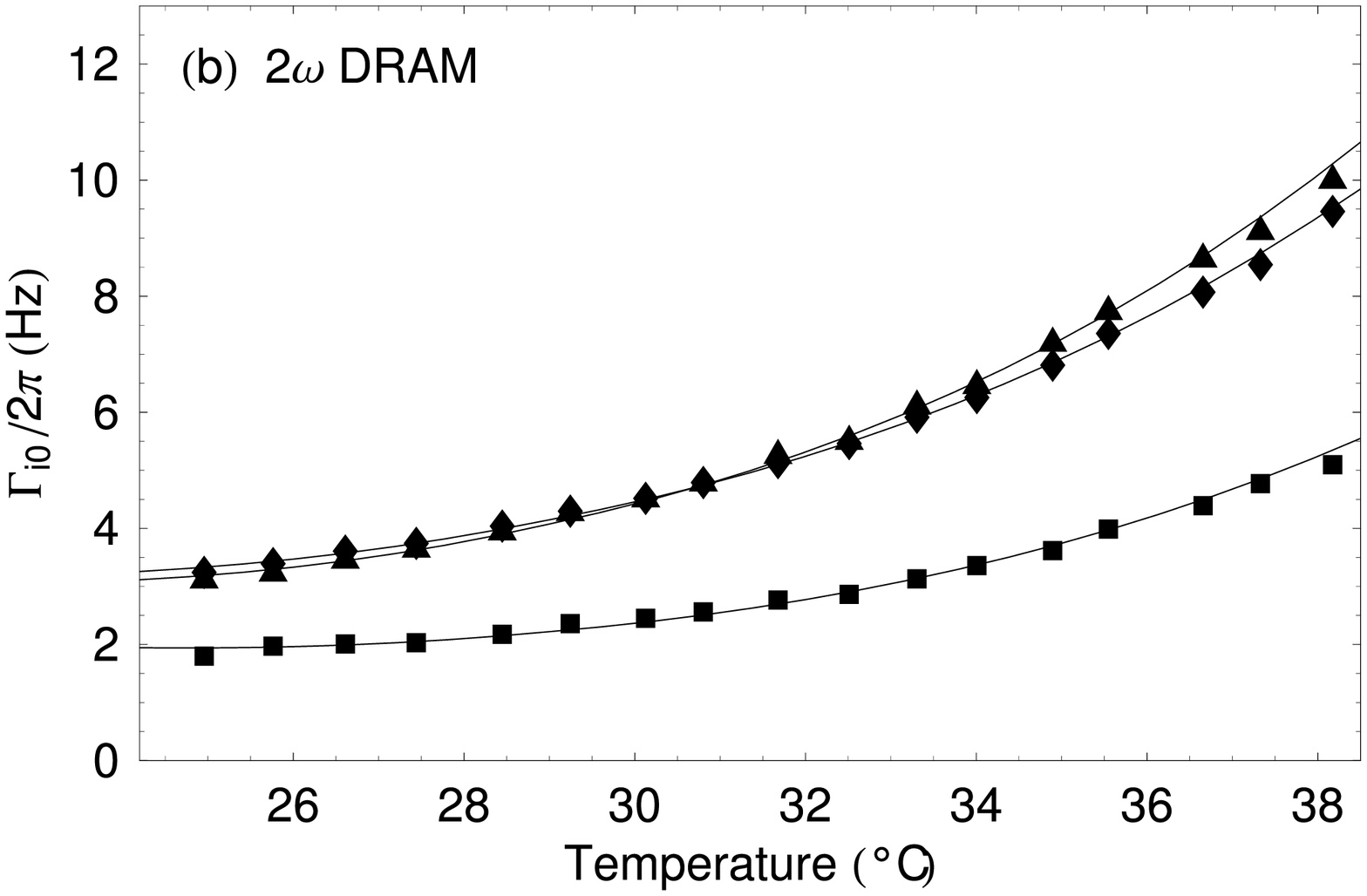}}
\caption{Temperature evolution of the relaxation rates extrapolated
to zero laser power. The symbols
($\blacksquare$,$\blacktriangle$,$\blacklozenge$) represent
$\sGamma{00}$, $\sGamma{10}$, $\sGamma{20}$. The solid lines are
fits of Eq.~(\ref{eq:relaxmodel}) to the experimental data. (a)
First harmonic DRAM with $\gamma=25.5\deg$. (b) Second harmonic DRAM
with $\gamma=90\deg$.} \label{fig:gammaT}
\end{figure}

The method developed in Sec.~\ref{sec:DRAMoptimum} --- predicting
the optimum point from a measurement of the extended DRAM model
parameters (cf.~Sec.~\ref{sec:DRAMextension}) --- was used to
investigate the temperature dependence of the optimum DRAM operating
point.  Multiple measurements of the DRAM parameters were made for
temperatures between $20^\circ$C and $40^\circ$C\@.  As an
illustration, the evolution of relaxation rates with temperature is
shown in Fig.~\ref{fig:GTPlot} where two measurements, at
$25^\circ$C and $38^\circ$C, are presented. Even though the
temperature evolution of the relaxation rates is non trivial, the
quadratic model of Eqs.~(\ref{eq:G1G2G3ext}) fits well to the
experimental data and gives access to the relaxation rates
$\sGamma{00}$, $\sGamma{10}$, and $\sGamma{20}$ at zero light power.
These parameters have been measured and their temperature behavior
is presented in Fig.~\ref{fig:gammaT}. They all increase with
temperature and this is mainly related to the increase of the atomic
vapor density. During setup, we observed that the difference between
$\sGamma{10}$ and $\sGamma{20}$ can be decreased by improving the
magnetic field homogeneity, and further tests confirmed that the
difference between $\sGamma{10}$ and $\sGamma{20}$ increases with
the square of the magnetic field
inhomogeneity~\cite{Abragam:1961:PNM,Pustelny:2006:IMF}.  Moreover,
the residual difference observed in our experiment is compatible
with an estimation of the residual magnetic field inhomogeneity.
This leads us to conclude that, in principle, the two transverse
relaxation rates $\sGamma{10}$ and $\sGamma{20}$ should be equal in
a perfectly homogeneous magnetic field.  In Fig.~\ref{fig:gammaT},
the solid lines are fits to the experimental data of the relaxation
model given by
\begin{equation}
    \sGamma{i0}=n\,\sigma_i\,v_\mathrm{rel}
    + A\,v_\mathrm{m}\,e^{E_a/kT}
    + B\,v_\mathrm{m}
    + C\,v_\mathrm{m}^{-1}.
    \label{eq:relaxmodel}
\end{equation}
On the right-hand side, the first term is the contribution due to
collisional spin-exchange, it is proportional to the vapor density
$n(T)$, to the collisional disalignment cross-section $\sigma_i$, and
to the atoms' mean relative velocity $v_\mathrm{rel}(T)=\sqrt{16 k
T/(\pi M)}$ where $M$ is the \magCs{} mass. The second term is the
contribution due to wall-collisions: it is proportional to the rate of
wall collisions, hence to the atoms' mean velocity
$v_\mathrm{m}(T)=\sqrt{8 k T/(\pi M)}$, and to the wall sticking time
$\tau_s\sim \tau_0 e^{E_a/kT}$ where $\tau_0\approx
10^{-12}~\mathrm{s}$, $E_a$ is the adsorption energy and $k$ is the
Boltzmann constant.  The third term is the contribution from the
reservoir effect, and it is proportional to the rate of wall
collisions and therefore to $v_\mathrm{m}$.  Finally, the last term is
the contribution due to magnetic field inhomogeneities, which is
proportional to $v_\mathrm{m}^{-1}$ due to motional
narrowing~\cite{Pustelny:2006:IMF}.

The vapor density $n(T)$ is calculated from the cesium vapor pressure
given in~\cite{TaylorLangmuir}: it is highly temperature dependent.
As a consequence, over the range of temperatures investigated in this
work, the collisional spin-exchange term represents the main
contribution to the temperature behavior of relaxation rates, and all
other terms are approximately linear.  Therefore, the fit is able to
determine the Cs--Cs collisions disalignment cross-sections
$\sigma_i$, but cannot distinguish the contributions from the other
terms with reasonable uncertainties.  In principle, this can be
improved by increasing the temperature range of the measurements, and
would lead to a powerful method for the investigation of relaxation
mechanisms.  However, at present, experimental setup cannot reach the
necessary temperatures and so, since it is beyond the scope of the
present paper, such investigations will be the subject of future work.
The values of $\sigma_i$ extracted from the fits are summarized in
Table~\ref{tab:relaxparameters}\@.

\begin{table}
\caption{Collisional disalignment cross-sections $\sigma_i$ obtained
from the fit of Eq.~(\ref{eq:relaxmodel}) to the experimental data
presented in Fig.~\ref{fig:gammaT}. The last column gives the
average of the values obtained for the $1\omega$ and $2\omega$
DRAMs.} \label{tab:relaxparameters} \centerline{
\begin{tabular}{|c|c|c|c|c|} \hline
\T Parameter  & $1\omega$ DRAM & $2\omega$ DRAM & Average
\\ \hline\hline
\T $\sigma_0$~($\mbox{cm}^2$) & $0.5(6)\times 10^{-14}$ &
$1.2(6)\times 10^{-14}$ & $0.9(4)\times 10^{-14}$\\
\T $\sigma_1$~($\mbox{cm}^2$) & $1.5(3)\times 10^{-14}$ &
$2.1(9)\times 10^{-14}$ & $1.6(3)\times 10^{-14}$\\
\T $\sigma_2$~($\mbox{cm}^2$) & $1.5(3)\times 10^{-14}$ &
$1.7(5)\times 10^{-14}$ & $1.6(3)\times 10^{-14}$\\
\hline
\end{tabular}
}
\end{table}

\begin{figure}[t]
\centerline{\includegraphics[width=\grscale\textwidth]{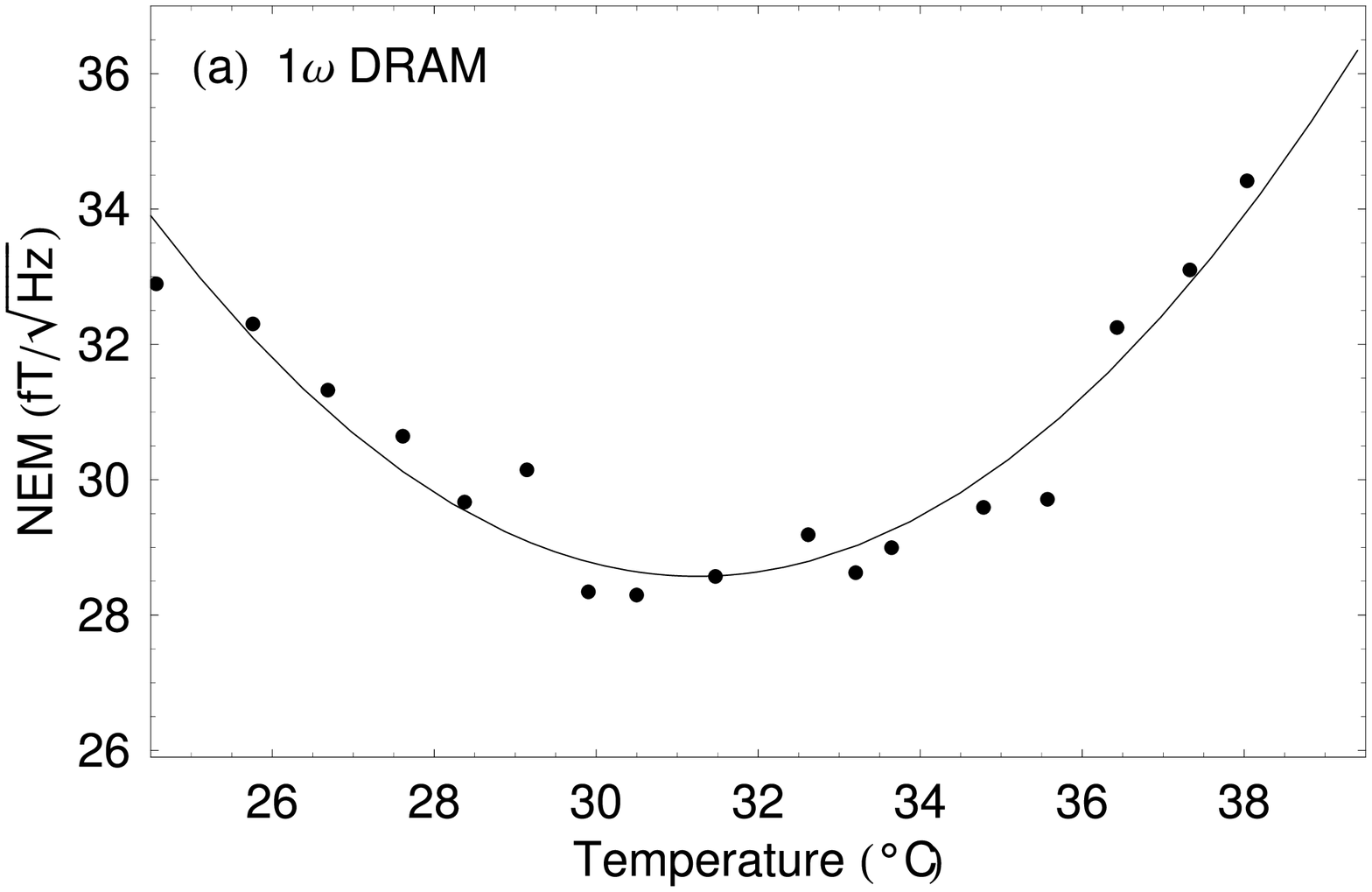}}
\centerline{\includegraphics[width=\grscale\textwidth]{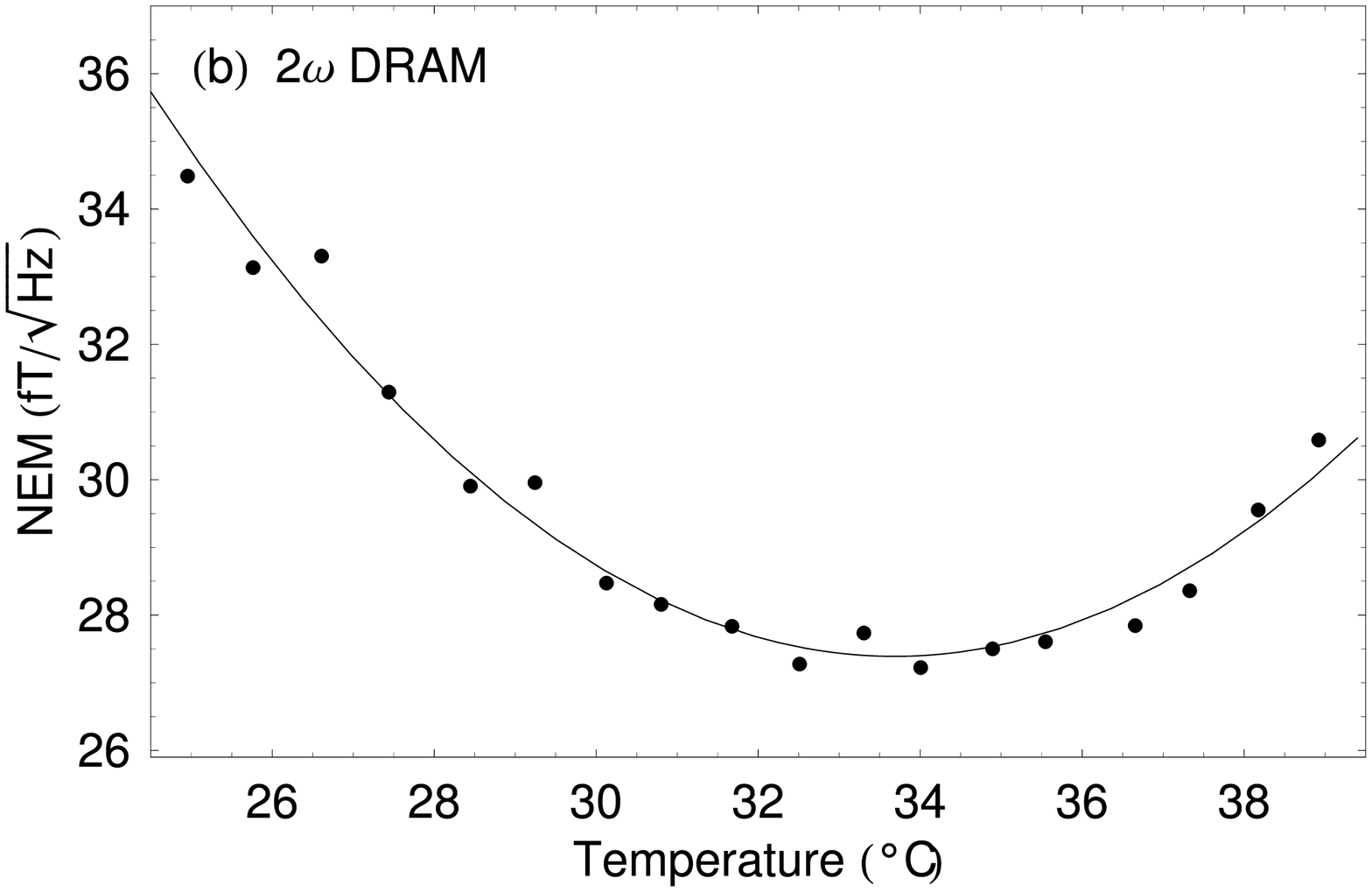}}
\caption{Evolution of the optimum \NEM{} as a function of
temperature. (a) First harmonic DRAM with $\gamma=25.5\deg$. (b)
Second harmonic DRAM with $\gamma=90\deg$. These graphs were
calculated from the measurement of the extended DRAM model
parameters.} \label{fig:optNEM}
\end{figure}

The extended DRAM model parameter measurements were used to calculate
the evolution of the optimum operating point of the DRAM
(cf.~Sec.~\ref{sec:DRAMoptimumtheo}).  The results are presented, as a
function of temperature, in Fig.~\ref{fig:optNEM} for both the first
and second harmonic DRAMs.  A quadratic polynomial was fitted to the
\NEM{} data in order to determine the temperature of minimum \NEM. The
experimental parameters characterizing these optimum points are
summarized in the first column of Table~\ref{tab:3cellsoptimum},
where we observe that the second harmonic DRAM is slightly more
sensitive than the first harmonic DRAM\@.

\section{Discussion}
\label{sec:discussion}

The majority of the work presented herein has been performed using a
single evacuated Cs cell (Cell~1 in Table~\ref{tab:3cellsoptimum}).
We applied the temperature \NEM{} optimization procedure described in
Sec.~\ref{sec:temperature} to two additional paraffin coated cesium
cells and the resulting optimum parameters are presented in
Table~\ref{tab:3cellsoptimum}\@.  The intrinsic \NEM{} for the second
harmonic DRAM is always smaller than for the first harmonic DRAM and
the lowest \NEM{} value of $27.4~\fTHz$ was obtained using the cell
produced by our group (Cell~1).  We observe that the \NEM{} values
scale with the inverse of the volume of the cell and not with the
volume to surface ratio.  However, since the three cells do not have
the same coating, this relation could be accidental.

Under identical conditions, the line shapes of the second harmonic
DRAM signal are narrower than those of the DROM which, in principle,
should lead to an improvement of the sensitivity~\cite{DRAMtheo}.
Previous work by our group found an intrinsic sensitivity of
$10~\fTHz$ for an optimized DROM using a 70~mm diameter Cs vapor cell
in the so called \Mx--configuration~\cite{FRAPLsOPM}. However, the
experimental setup used in the past was very different (different cell
size, offset field homogeneity, and magnetic shielding) and therefore
a detailed comparative study is needed before drawing firm
conclusions.

\begin{table}
\caption{Results of the temperature \NEM{} optimization procedure (as
described in Sec.~\ref{sec:temperature}) applied to three different
paraffin coated cells.  Cell~1 was produced by our group, it is
spherical with a 28~mm diameter.  Cells~2 and~3 were purchased from
a Russian company.}
\label{tab:3cellsoptimum}
\centerline{
\begin{tabular}{|l|c|c|c|}
\hline
                        & \textbf{Cell 1 }  & \textbf{Cell 2  } & \textbf{Cell 3 }      \\ \hline
          Shape         & Spherical         & Cubic             & Cylindrical           \\
          Volume $V$    & $11.5~\mbox{cm}^3$& $8.0~\mbox{cm}^3$ & $4.6~\mbox{cm}^3$     \\
          Surface $S$   & $24.6~\mbox{cm}^2$& $24.0~\mbox{cm}^2$& $15.3~\mbox{cm}^2$    \\
          $V/S$         & $0.467~\mbox{cm}$ & $0.333~\mbox{cm}$ & $0.301~\mbox{cm}$     \\
\hline
\multicolumn{4}{|c|}{\textbf{First harmonic DRAM}} \\
\hline
Optimum T               & $31.2~{}^\circ$C  & $35.8~{}^\circ$C  & $35.5~{}^\circ$C      \\
Optimum $\PL$           & $6.2~\uW$         & $12.1~\uW$        & $5.2~\uW$             \\
Optimum $\omega_1/2\pi$ & $2.9~\Hz$         & $5.5~\Hz$         & $4.8~\Hz$             \\
Intrinsic \NEM{}        & $28.6~\fTHz$        & $45.7~\fTHz$        & $74.7~\fTHz$      \\
\hline
\multicolumn{4}{|c|}{\textbf{Second harmonic DRAM}}\\
\hline
Optimum T               & $33.7~{}^\circ$C  & $36.2~{}^\circ$C  & $37.6~{}^\circ$C      \\
Optimum $\PL$           & $6.0~\uW$         & $8.9~\uW$         & $4.7~\uW$             \\
Optimum $\omega_1/2\pi$ & $10.5~\Hz$        & $14.1~\Hz$        & $15.0~\Hz$            \\
Intrinsic \NEM{}        & $27.4~\fTHz$        & $39.3~\fTHz$        & $62.2~\fTHz$      \\
\hline
\end{tabular}
}
\end{table}

\section{Conclusion}

In conclusion, we have presented an experimental study of the
intrinsic magnetometric sensitivity of the double-resonance
alignment magnetometer, showing that an empirical extension of the
DRAM model can be used to describe the magnetic resonance spectra
over a range of experimental parameters sufficient for optimizing
the magnetometer.  A model has been developed to predict the optimum
operating point of the magnetometer, i.e., the value of experimental
parameters for which the magnetometric sensitivity is maximum.  The
method was verified by comparing its results to a direct measurement
of the optimum operating point.  In contrast to the time consuming
direct optimization involving many hours of testing in a two
parameter space, our method decreases the time required to find the
optimum operation point to half an hour.  Finally, we used this
method to investigate the evolution of the optimum operating point
of the DRAM with temperature, showing that the magnetometric
sensitivity reaches an optimum of $27.4\:\fTHz$ for a temperature of
$33.7^\circ$C\@.  Both the first harmonic and the second harmonic
realizations of the magnetometer were explored and compared. The
temperature dependence of the relaxation rates yielded measurements
of the Cs--Cs collisional disalignment cross sections of the tensor
alignment, and the method promises to be useful in the continued
study of the relaxation processes over broader temperature ranges.

\begin{acknowledgments}
This work was supported by grants from the Swiss National Science
Foundation (Nr.~205321--105597, 200020--111958), from the Swiss
Innovation Promotion Agency, CTI (Nr.~8057.1 LSPP--LS), from the
Swiss Heart Foundation, and from the Velux foundation.
\end{acknowledgments}


\end{document}